\documentclass[aps,prl,reprint,10pt,nofootinbib]{revtex4-2}
\usepackage[T1]{fontenc}
\usepackage{newtxtext,newtxmath}
\usepackage{amsmath,bm,graphicx,booktabs,microtype}
\graphicspath{{./}{figures/}{../figures/}}
\usepackage[colorlinks=true,allcolors=black]{hyperref}
\hypersetup{pdfauthor={Itai Panas},pdftitle={Contact Renormalization of One-Mode Electron Correlation: Validation in Four- and Ten-Electron Systems}}
\newcommand{\Eh}{E_{\mathrm h}}
\newcommand{\Ec}{E_{\mathrm c}}
\newcommand{\Mvv}{M_0^{VV}}
\newcommand{\Dvv}{D_{\mathrm{eff}}^{VV}}
\newcommand{\TV}{T_V}

\newcommand{\nV}{n_V}
\newcommand{\gV}{g_V}
\begin{document}
\title{Contact Renormalization of One-Mode Electron Correlation: Validation in Four- and Ten-Electron Systems}
\author{Itai Panas}
\affiliation{Department of Chemistry and Chemical Engineering, Chalmers University of Technology, SE-412 96 Gothenburg, Sweden}
\date{22 October 2026}
\begin{abstract}
A stationary many-electron eigenstate has one dynamical phase although its internal response can be strongly multimode. We ask whether the stationary Hartree--Fock information carrier contains enough information to determine the scalar correlation energy through one effective return. In regularized Hartree--Fock, Gaussian primitive products fix a minimum-information relative-coordinate mode before any one-mode reduction; virial completion defines the correlation source, while exact contact measures the additional return restriction created by neighboring occupied pairs. The Be-like $1s^22s^2$ sequence establishes the one-valence-pair boundary: with no neighboring occupied valence pair, the source returns through the universal floor $D_0=\sqrt{1+\delta^2}$. Ten-electron molecules and Ne-like ions then test how finite-pair contact dresses that return. The same parameter-free source--return construction reproduces the established molecular and ionic correlation trends with quantitative accuracy. The numerical agreement therefore serves as validation of a physical separation between source strength and return restriction, rather than as the objective by itself.
\end{abstract}
\maketitle

A stationary many-particle eigenstate has one temporal phase, irrespective of the number of configurations needed to describe its internal correlations:
\begin{equation}
|\Psi(t)\rangle=e^{-iEt/\hbar}\sum_I c_I|\Phi_I\rangle.
\label{eq:phase}
\end{equation}
All configurations share the same eigenfrequency; their amplitudes, relative phases, and entanglement need not be simple. This common phase does not imply wave-function factorization, a single excitation pole, or the exactness of Hartree--Fock. It motivates a different question: how much information is needed to predict the \emph{one scalar energy} that generates stationary time evolution?

The preceding one-mode construction\cite{Panas} separates exact scalar compression from its predictive realization and, crucially, specifies the information carrier \emph{before} reducing the return. Canonical restricted Hartree--Fock (RHF) Gaussian primitive products fix a relative-coordinate Gaussian measure and its covariance. At fixed covariance that same Gaussian minimizes Fisher information, maximizes Shannon differential entropy, and saturates the Stam bound\cite{Panas}. These are not three extra assumptions: they identify the same minimum-information pair mode already fixed by the stationary HF representation. The mean field is therefore used as an information carrier for the scalar energy, not as an approximation to the detailed correlated wave function.

Virial completion then converts the scale-resolved Coulomb displacement into the stationary-compatible reg-HF response. The one-mode exclusion rule of the preceding work excludes covariance carried by response components orthogonal to the retained pair mode from the \emph{terminal scalar energy}; it does not assert that those components are absent or that the microscopic response has one pole\cite{Panas}. The correlation source and return are constructed downstream, without feeding the final correction back into the determinant. The Be-like $1s^22s^2$ sequence tests the one-pair boundary; CH$_4$--Ne and Ne--Ar$^{8+}$ test finite-pair contact control. Jarz\k{e}cki and Davidson emphasized the strongly $Z$-dependent Be-like correlation and contrasting Ne-like behavior\cite{JD98}; numerical comparisons use the refined Chakravorty--Davidson estimates\cite{CD96}.

\emph{Information carrier, source, and one-mode return.---}
For a specified reference, write the source as $|S\rangle$ and the complementary-space operator at fixed starting energy as $L+U$, with $L$ positive and invertible on the source-supported subspace. Set $|F\rangle=L^{-1}|S\rangle$, $A=\langle S|F\rangle$, and $B=\langle F|U|F\rangle$. Projection onto the direction $|F\rangle$ gives
\begin{equation}
\Delta_{1\mathrm m}=-\frac{A^2}{A+B}=-\frac{A}{1+B/A}.
\label{eq:onemode}
\end{equation}
This is a declared one-direction, fixed-energy approximation, not a consequence of Eq.~(\ref{eq:phase}) alone. A broad microscopic return can contribute to its scalar matrix elements. If the reference is not RHF, the conventional correlation increment requires $\Ec=\Delta_{1\mathrm m}-(E_{\rm HF}-E_{\rm ref})$. The present energies are defined directly relative to stationary RHF; the Supplemental Material separates exact projection, scalar compression, and this predictive closure\cite{SM}.

The bare source is retained in the notation and production convention of the preceding reg-HF work:
\begin{equation}
I_{ij}^{\rm bare}=\left|2J_{ij}^{\rm corr}-K_{ij}^{\rm corr}\right|,
\qquad
\Mvv=\sum_{i\leq j\in V} I_{ij}^{\rm bare}.
\label{eq:source}
\end{equation}
Here ``corr'' denotes the established virial-completed reg-HF four-index source tensor\cite{Panas,SM}; it is not the native Gaussian contact kernel and not the ordinary bare-Coulomb ERI tensor. The canonical occupied orbitals and positive unique-pair summation are kept fixed. Thus the present work changes the return, not the upstream source construction.

The entire test changes only the VV denominator:
\begin{equation}
\Ec=-CC^+-CV^+-\frac{\Mvv}{\Dvv}.
\label{eq:energy}
\end{equation}
The original positive core--core and core--valence contributions retain their common local-$Z$ confinement. This shared return scale is robust to basis-dependent redistribution between the CC and CV bookkeeping sectors: orthogonalization can move core/valence character between occupied orbitals without assigning a different return price to that relabeling. We therefore regard $CC^++CV^+$ as one core sector. No additional core contact dressing is introduced, isolating the consequences of the proposed VV closure.

\emph{An exact HF contact and a finite-pair closure.---}
For $N_V$ doubly occupied valence spatial orbitals, define $\nV=\sum_{i\in V}|\phi_i|^2$, with $\int\nV=N_V$. This is a one-spin density; the spin-summed density is $2\nV$. For a three-dimensional Dirac kernel, direct and exchange matrix elements coincide at electron coalescence. Complete ordered valence contraction therefore gives
\begin{equation}
\TV=2J_V[\delta^{(3)}]-K_V[\delta^{(3)}]
=\int\nV^2(\mathbf r)\,d^3r.
\label{eq:contact}
\end{equation}
The same-spin contact cancels, and the opposite-spin HF on-top contact survives. Equation~(\ref{eq:contact}) is exact and invariant under occupied rotations within the valence subspace. Its conversion into a return penalty is a separate model assumption: an opposite-spin contact is not itself a same-spin repulsive potential.

A finite-pair Gross--Pitaevskii-type (GP) coordinate\cite{Gross,Pitaevskii} is
\begin{equation}
\kappa_0=\gV a_0^3\TV,\qquad
\gV=\frac{N_V-1}{2N_V}.
\label{eq:gp}
\end{equation}
For $\Psi_V=\sqrt{\nV/N_V}$ this equals
$\binom{N_V}{2}a_0^3\int|\Psi_V|^4d^3r$.
The analogy concerns pair counting, not bosonic electron statistics. Indeed, writing $p_i=|\phi_i|^2$, one has the exact bound
\begin{equation}
\sum_{i<j}\int p_ip_j\,d^3r
=\frac12\left(\TV-\sum_i\int p_i^2d^3r\right)
\leq\gV\TV.
\label{eq:bound}
\end{equation}
Thus Eq.~(\ref{eq:gp}) is an equal-sharing upper envelope for distinct-orbital contact. It models the maximal similarity-driven restriction on a collective return, rather than calculating a correlated Pauli projector.

We use the established shifted-square-root law
\begin{equation}
\Dvv=\sqrt{(1+x_V)^2+\delta^2},\qquad
\delta=\frac{7\sqrt5-15}{4},
\label{eq:denom}
\end{equation}
with $x_V=\kappa_0$ for the neutral systems. It preserves $D_0=\sqrt{1+\delta^2}$ when $N_V=1$, where no additional interpair restriction is present. The ten-electron systems have $N_V=4$ and hence $\gV=3/8$; the Be-like floor test has $N_V=1$ and $\gV=0$. Equation~(\ref{eq:denom}) is reconciled with the one-mode form, Eq.~(\ref{eq:onemode}), by defining the one-pair reference amplitude $A_0=\Mvv/D_0$: $-\Mvv/\Dvv=-A_0/(\Dvv/D_0)$. Under the explicit matching $A=A_0$, Eq.~(\ref{eq:onemode}) gives $1+B/A=\Dvv/D_0$, hence $B/A=\Dvv/D_0-1$. Thus the projected scalar return sees only the norm ratio: $x_V$ first shifts the active stiffness $Q_V=1+x_V$ and is not itself $B/A$. The matching is a closure hypothesis, not a consequence of the contact identity.

The square-root return has a useful long-range analogue. In a two-dimensional electron liquid, Yukawa regularization gives $v_Q(\kappa)=2\pi e^2/\sqrt{Q^2+\kappa^2}$\cite{GiulianiVignale}. Here the correspondence is metric, not a claim of physical two-dimensionality. For a general resolved stiffness $Q$, let $D(Q)=\sqrt{Q^2+\delta^2}$; the production valence branch is $Q_V=1+x_V$, so $\Dvv=D(Q_V)$. One active coordinate is therefore embedded with the fixed orthogonal Gaussian-information invariant $\delta$. The susceptibility $\chi(Q)=-1/D(Q)$ remains finite in the zero-confinement continuation, $\chi(0)=-1/\delta$; the Coulomb source itself then vanishes, so no correlation survives without interaction. Unlike the auxiliary Yukawa regulator, $\delta$ is intrinsic to the Gaussian Fisher--Shannon--Stam closure.

\begin{figure*}[t]
\centering
\includegraphics[width=\textwidth]{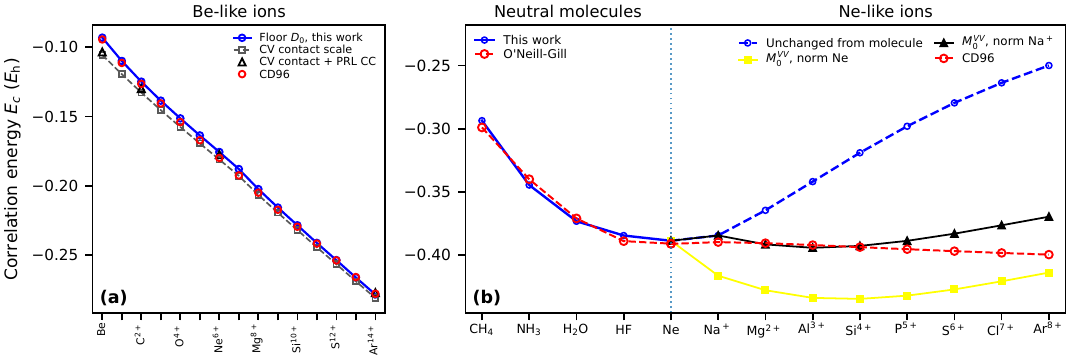}
\caption{(a) Be-like $1s^22s^2$ core-sector audit at RHF/6-311G(s-only): adopted common local-$Z$ $CC+CV$ plus the VV floor (blue), CD96 totals (red), CV contact-scale control (gray), and the same control with first-PRL CC return at four tabulated checkpoints (black triangles, no interpolation). The controls are diagnostic only. (b) Retained ten-electron comparison: CH$_4$--HF, Ne, and the Ne-like ionic sequence with the three unchanged theory curves; references are O'Neill--Gill\cite{ONeillGill} and CD96\cite{CD96}.}
\label{fig:ions}
\end{figure*}

\emph{One-pair floor.---}
For Be-like $1s^22s^2$, $1s^2$ remains core and $2s^2$ is the sole VV pair, so $x_V=0$ and
\begin{equation}
M_{-1}^{VV}=\frac{\Mvv}{D_0},\qquad D_0=\sqrt{1+\delta^2}=1.01321655894.
\label{eq:floor}
\end{equation}
A complete RHF/6-311G(s-only) $Z=4$--18 sequence has $1.87$ m$\Eh$ mean and $4.63$ m$\Eh$ maximum absolute deviation from the $N=4$ Chakravorty--Davidson estimates; Table~\ref{tab:ions} shows representative points and the Supplemental Material gives all 15\cite{SM}. Because Eq.~(\ref{eq:floor}) fixes VV, the totals conditionally define $C_{\rm req}^+=-E_c^{\rm CD}-M_{-1}^{VV}$ and therefore test the combined $C^+=CC^++CV^+$ sector, not CC and CV separately. Figure~\ref{fig:ions}(a) restores two no-fit core controls; the direct $C^+$ audit and repartitioning argument are in the Supplemental Material\cite{SM}.

\begin{table*}[t]
\caption{Neutral ten-electron systems. The fresh basis $b_1$ is the project-specific 6-311G(best,5D,p): radially augmented s/p functions, five traceless d functions on the heavy atom, and p polarization on H. The archived baseline $b_0$ is 6-311G(d,p). Energies are in hartree and $T_V$ in $a_0^{-3}$. $\Delta_1=10^3(E_c^{(b_1)}-E_c^{\rm ref})$ is in m$E_{\rm h}$. All five $E_c^{\rm ref}$ values are the RHF correlation estimates of O'Neill--Gill, Table 2\cite{ONeillGill}, printed as integer m$E_{\rm h}$ and shown here to three decimals. Deviations are rounded to $0.1$ m$E_{\rm h}$ for descriptive comparison, not sub-millihartree reference accuracy. Table~\ref{tab:ions} uses a separate atomic reference set. Baseline contacts used numerical grids; fresh contacts are analytic. Geometry and source conventions are given in the Supplemental Material.}
\label{tab:neutral}
\begin{ruledtabular}
\begin{tabular*}{\textwidth}{@{\extracolsep{\fill}}lrrrrrr}
System & $E_{\rm HF}^{(b_1)}$ & $T_V^{(b_1)}$ & $E_c^{(b_0)}$ & $E_c^{(b_1)}$ & $E_c^{\rm ref}$ & $\Delta_1$\\
\hline
CH$_4$ & $-40.210493$ & 0.253857 & $-0.29327$ & $-0.293640$ & $-0.299$ & $+5.4$ \\
NH$_3$ & $-56.215562$ & 0.410509 & $-0.34632$ & $-0.344584$ & $-0.340$ & $-4.6$ \\
H$_2$O & $-76.055301$ & 0.720623 & $-0.37417$ & $-0.373265$ & $-0.371$ & $-2.3$ \\
HF & $-100.058314$ & 1.277345 & $-0.38568$ & $-0.384565$ & $-0.389$ & $+4.4$ \\
Ne & $-128.541115$ & 2.193704 & $-0.39216$ & $-0.388726$ & $-0.391$ & $+2.3$ \\
\end{tabular*}
\end{ruledtabular}
\end{table*}

\emph{Neutral series.---}
Table~\ref{tab:neutral} reports fresh RHF and source calculations. Across CH$_4$--Ne, $T_V$ rises from $0.25386$ to $2.19370\,a_0^{-3}$, while $E_c$ changes from $-0.29364$ to $-0.38873\,\Eh$. The mean and maximum absolute deviations from O'Neill--Gill's printed RHF correlation estimates\cite{ONeillGill} are approximately $3.8$ and $5.4$ m$\Eh$. Reference precision remains one millihartree, including Ne at $-0.391\,\Eh$. Fixed catalog geometries are not a verified reproduction of every reference geometry, so this is not a geometry-matched accuracy claim. No target was inverted to determine a denominator.

\emph{Source-relative contact growth.---}
An ionic contraction increases the probability of exact coincidence without necessarily creating the same proportional increase in collective return restriction. The Dirac contact is zero range, whereas the inherited virial-completed reg-HF source is finite range in the Gaussian representation. The source $\Mvv$ therefore supplies a finite-range measure with which to normalize the zero-range contact. This use preserves the canonical production convention of Eq.~(\ref{eq:source}); it does not replace the source by an ordered signed expectation.

For the ionic sequence, at a fixed basis $b$, define
\begin{equation}
x_0(Z;b)=\tfrac38 a_0^3\TV(Z;b),\qquad
x_a(Z;b)=x_0(Z;b)\,\mathcal R_a(Z;b),
\label{eq:normalizations}
\end{equation}
with reference $a=10$ (Ne) or $a=11$ (Na$^+$), and
\begin{equation}
\mathcal R_a(Z;b)=
\begin{cases}
1,&10\leq Z\leq a,\\[3pt]
\Mvv(a;b)/\Mvv(Z;b),&Z>a.
\end{cases}
\label{eq:onset}
\end{equation}
The three comparisons are thus \emph{unchanged}, \emph{$M_0^{VV}$ reference Ne}, and \emph{$M_0^{VV}$ reference Na$^+$}. Neutral molecules retain $x_V=\kappa_0$ in every case. Ne is unchanged for both references; the Na$^+$-referenced law also leaves Na$^+$ at its unchanged-GP value. Its renormalization applies only beyond $Z=11$. Numerator and denominator of each source ratio use the same basis and source convention.

For $Z\geq a$, the normalized contact obeys
\begin{equation}
\frac{\widetilde T_V^{(a)}(Z)}{\TV(a)}
=\frac{\TV(Z)/\TV(a)}{\Mvv(Z)/\Mvv(a)}.
\label{eq:differential}
\end{equation}
It measures zero-range contact growth relative to Gaussian-tailed source growth. The ratio is dimensionless and supplies no correlated energy. Only the contact coordinate is normalized; the autonomous unit term in $1+x_V$, the source numerator, and CC/CV are unaltered.

\emph{Radial confinement and angular return.---}
We interpret the Ne--Na$^+$ offset as a possible change in return character. The unchanged closure is hypothesized to remain adequate through Na$^+$ for a predominantly radial response, while coherent angular reconfiguration of $2p^4ab$ double excitations can redistribute the source-supported return at higher charge. Such configurations already exist in Ne; no new hole count or ground-state shell transition is implied. ``Asymmetric'' means departure from an autonomous source-aligned return, not loss of Hermiticity. Neither radial dominance nor a crossover at $Z=11$ is established by a channel-resolved calculation; the reference choice remains a tested closure hypothesis.

\begin{table*}[t]
\caption{Representative atomic points. Be-like rows use the one-pair floor at RHF/6-311G(s-only); Ne-like rows retain RHF/6-311G(best) and the three established denominators. $q=Z-N_e$. $E_c^{\rm base}$ is the Be-like floor or the Ne-like unchanged result; $E_c^{10}$ and $E_c^{11}$ are the Ne- and Na$^+$-referenced Ne-like values. $E_c^{\rm CD}$ is Chakravorty--Davidson\cite{CD96}; $\Delta$ uses the Be-like floor or the Na$^+$-referenced Ne-like value. The complete 15- and 9-ion tables are in the Supplemental Material. Energies are in hartree and $\Delta$ in m$E_{\rm h}$.}
\label{tab:ions}
\begin{ruledtabular}
\begin{tabular*}{\textwidth}{@{\extracolsep{\fill}}crlrrrrr}
$N_e$ & $q$ & Ion & $E_c^{\rm base}$ & $E_c^{10}$ & $E_c^{11}$ & $E_c^{\rm CD}$ & $\Delta$\\
\hline
4 & 0 & Be & $-0.093054$ & --- & --- & $-0.09436$ & $+1.31$ \\
4 & 6 & Ne$^{6+}$ & $-0.175627$ & --- & --- & $-0.17995$ & $+4.32$ \\
4 & 14 & Ar$^{14+}$ & $-0.278524$ & --- & --- & $-0.27800$ & $-0.52$ \\
\hline
10 & 0 & Ne & $-0.388724$ & $-0.388724$ & $-0.388724$ & $-0.39120$ & $+2.48$ \\
10 & 1 & Na$^+$ & $-0.384520$ & $-0.416386$ & $-0.384520$ & $-0.38978$ & $+5.26$ \\
10 & 2 & Mg$^{2+}$ & $-0.364609$ & $-0.427886$ & $-0.391584$ & $-0.39060$ & $-0.98$ \\
10 & 4 & Si$^{4+}$ & $-0.319046$ & $-0.434724$ & $-0.392781$ & $-0.39377$ & $+0.99$ \\
10 & 8 & Ar$^{8+}$ & $-0.249967$ & $-0.413908$ & $-0.369660$ & $-0.39968$ & $+30.02$ \\
\end{tabular*}
\end{ruledtabular}
\end{table*}

\emph{Many-pair isoelectronic test.---}
Representative Ne-like results are in Table~\ref{tab:ions}; the complete nine-ion sequence is in the Supplemental Material and panel (b) of Fig.~\ref{fig:ions}. From Ne to Ar$^{8+}$, $T_V$ increases by a factor of $13.29$, while $\Mvv$ increases by $2.79$. The unchanged model changes from $-0.38872$ to $-0.24997\,\Eh$, suppressing much of the VV contribution. Referencing the source ratio to Ne reduces the full nine-ion span from $138.76$ to $46.00$ m$\Eh$; onset at Na$^+$ reduces it to $24.63$ m$\Eh$.

The origin of the difference is transparent: $\Mvv(11)/\Mvv(10)=1.23656$. For $Z\geq11$, the Na$^+$-referenced contact coordinate is therefore $23.66\%$ larger than the Ne-referenced one. The stronger denominator removes much of the intermediate-ion minimum but produces a less-negative high-charge tail. The smaller complete-sequence span is not a uniform improvement in flatness: for the charged ions alone ($Z=11$--18), the Ne- and Na$^+$-referenced spans are $20.82$ and $24.63$ m$\Eh$.

Chakravorty and Davidson combined corrected experimental ionization potentials, multiconfiguration reference calculations, high-$Z$ constraints, and interpolation to obtain refined nonrelativistic estimates\cite{CD96}. Their nine-ion span is $9.90$ m$\Eh$. Mean absolute deviations are $73.09$, $28.10$, and $9.36$ m$\Eh$ for unchanged GP, reference Ne, and reference Na$^+$, respectively. The Na$^+$ trial lies within $2.19$ m$\Eh$ of their estimates for Mg$^{2+}$--Si$^{4+}$, but becomes progressively less negative at larger $Z$, reaching a $+30.02$ m$\Eh$ deviation at Ar$^{8+}$.

These are constrained reference estimates, not direct measurements of correlation energy or exact finite-basis CI results. Conversely, absence of energy fitting does not establish the accuracy of a closure. We retain the predictions and their disagreements. Since Eq.~(\ref{eq:energy}) includes inherited CC and CV terms, the total correlation-energy comparison cannot assign all disagreement to VV dressing.

\emph{Numerical control and scope.---}
Contacts are analytic. Fresh neutral sources use pure-5D polarization, orthonormal-density convergence, and projected Fock commutator norms $0.89$--$2.32\times10^{-8}$. One near-dependent overlap direction is removed for NH$_3$, H$_2$O, and Ne. Earlier anomalous Cartesian-6D/source-cache results are excluded. The ionic ``best'' basis is the documented s/p-only radial augmentation, not the neutral polarized basis.

For Ne, Na$^+$, Mg$^{2+}$, and Ar$^{8+}$, the Ne-source-referenced spans are $35.47$, $38.56$, and $39.16$ m$\Eh$ in 6-31G, 6-311G, and 6-311G(best). Best-minus-6-311G changes are at most $1.02$ m$\Eh$. This establishes stability under the tested enlargement, not complete-basis convergence or a separate multibasis test of the Na$^+$ onset. The source remains unchanged; unresolved correspondence between native Gaussian auxiliaries and the production representation is documented separately\cite{SM}.

The tests cover the Be-like one-pair boundary for $4\le Z\le18$ and the ten-electron family for $10\le Z\le18$. The Be-like production sequence is uniformly 6-311G(s-only); the incomplete radially augmented control and its conditioning qualification are documented in the Supplemental Material\cite{SM}. The results do not identify exact $B/A$, establish size consistency, or fix a universal high-density limit. Residual curvature is retained rather than removed by an additional cutoff or charge fit.

\emph{Conclusion.---}
One stationary phase is compatible with a multimode internal response. The hydrogenic one-electron Schrödinger form, generalized variationally to coupled HF equations, supplies the stationary information carrier, while the Gaussian Heisenberg--Fisher--Shannon--Stam closure fixes the invariant $\delta$. Thus $D^2=q^2+\delta^2$ is formally two-component but one-mode: $\delta$ is fixed and only $q$ is elastic. H$^-$ already exhibits this elasticity, with $q=0.86858<Z=1$ under self-consistent screening\cite{Panas}, while strong Coulomb confinement drives the resolved scale toward $Z$. The same square-root architecture occurs in the Yukawa-regularized 2D electron-liquid kernel, without implying physical two-dimensionality or a one-pole microscopic response. All calculations ran on an entry-level three-year-old laptop; no energy-use benchmark is claimed.

\bibliographystyle{apsrev4-2}
\bibliography{references}
\end{document}